\documentclass[aps,10pt]{revtex4}

\usepackage{amsmath}
\usepackage{graphicx}

\begin{document}

\title{
Phase reduction approach to synchronization of
spatiotemporal rhythms in reaction-diffusion systems
}

\author{Hiroya Nakao}
\email{nakao@mei.titech.ac.jp}
\affiliation{Graduate School of Information Science and Engineering, Tokyo Institute of Technology, Tokyo 152-8552, Japan}

\author{Tatsuo Yanagita}
\affiliation{Osaka Electro-Communication University, Neyagawa 572-8530, Japan}

\author{Yoji Kawamura}
\affiliation{Department of Mathematical Science and Advanced Technology,
Japan Agency for Marine-Earth Science and Technology, Yokohama 236-0001, Japan}

\date{\today}

\begin{abstract}
Reaction-diffusion systems can describe a wide class of rhythmic spatiotemporal patterns observed in chemical and biological systems, such as circulating pulses on a ring, oscillating spots, target waves, and rotating spirals.
These rhythmic dynamics can be considered limit cycles of reaction-diffusion systems.
However, the conventional phase-reduction theory, which provides a simple unified framework for analyzing synchronization properties of limit-cycle oscillators subjected to weak forcing,
has mostly been restricted to low-dimensional dynamical systems.
Here, we develop a phase-reduction theory for stable limit-cycle solutions of infinite-dimensional reaction-diffusion systems.
By generalizing the notion of isochrons
to functional space, the phase sensitivity function --- a fundamental quantity for phase reduction --- is derived.
For illustration, several rhythmic dynamics of the FitzHugh-Nagumo model of excitable media are considered.
Nontrivial phase response properties and synchronization dynamics are revealed,
reflecting their complex spatiotemporal organization.
Our theory will provide a general basis for the analysis and control of spatiotemporal rhythms in various reaction-diffusion systems.
\end{abstract}

\maketitle

\section{Introduction}

The phase-reduction theory provides a general framework to simplify multidimensional ordinary differential equations (ODEs) describing weakly perturbed limit-cycle oscillators
to one-dimensional approximate phase equations~\cite{winfree80,kuramoto,hoppensteadt97,brown04,ermentrout10}.
It has drastically facilitated theoretical and experimental analysis of the synchronization properties of weakly interacting nonlinear oscillators such as chemical oscillators and spiking neurons~\cite{winfree80,kuramoto,hoppensteadt97,brown04,ermentrout10,pikovsky01,hudson,taylor,tinsley}.
Methods for controlling limit-cycle oscillators have also been developed on the basis of the phase reduction theory~\cite{kiss,moehlis,harada,zlotnik}.

In real-world systems, rhythmic dynamics often arise collectively from a number of spatially distributed interacting elements, rather than from a single isolated oscillator, e.g., heartbeats generated by an ensemble of pulsating cardiac cells~\cite{winfree80,glass0,glass1,bub}.
Such systems are often modeled by reaction-diffusion (RD) systems, and the collective spatiotemporal rhythms are described by stable limit-cycle solutions of the RD systems~\cite{winfree80,kuramoto,rabinovich,mikhailov,hastings,hagberg,nomura,yanagita,ertl,tyson,maneville}.
Synchronization of collective spatiotemporal rhythms has been investigated experimentally in chemical systems~\cite{hildebrand01,fukushima} and may be of significant practical importance, e.g., in biomedical engineering~\cite{glass0,glass1,bub}.
In order to analyze and control the dynamics of collective spatiotemporal rhythms, it is desirable to develop a phase-reduction theory for the RD systems.

Various types of low-dimensional phase equations have been derived for RD systems, in particular for traveling pulses~\cite{kuramoto,tyson,ei,maneville,cross,mori,ohta,ermentrout,lober} and for rotating spirals~\cite{sandstede,biktashev}.
In most cases, however, it is assumed that the system is symmetric with respect to continuous spatial translation or rotation and the spatial structure is rigidly translating or rotating, so that their location or angle is simply identified as the phase.
However, such assumptions exclude various intriguing rhythmic dynamics of RD systems that lack continuous spatial symmetry.
Since limit cycles are essentially associated with {\em temporal} translational symmetry, we should be able to derive phase equations from general RD systems without recourse to spatial symmetry.

Our goal in the present study is to develop, without assuming any spatial symmetry or rigidity, a phase-reduction theory for weakly perturbed RD systems exhibiting stable rhythmic dynamics.
We solve this problem by generalizing the conventional phase-reduction theory for ODEs to RD systems.
Our theory gives a systematic method to approximate rhythmic dynamics of infinite-dimensional RD systems by one-dimensional phase equations, thereby facilitating detailed analysis of the synchronization dynamics of rhythmic spatiotemporal patterns.
As a simple example, we analyze mutual synchronization between two interacting layers of RD systems exhibiting rhythmic dynamics.
The proposed theory provides a simple, unified description of rhythmic spatiotemporal patterns and will be the basis for developing methods to control and design rhythmic spatiotemporal patterns in RD systems.

\section{Phase description of spatiotemporal rhythms}

In this section, we summarize the essential results of the proposed phase reduction theory for RD systems and apply it to mutual synchronization of a pair of coupled RD systems. Full derivation of the theory will be given in Appendix B.  See also Appendix~A for a review of the phase reduction theory for ordinary limit-cycle oscillators described by ODEs.

\subsection{Phase reduction of limit-cycle solutions in reaction-diffusion systems}

\begin{figure}[htbp]
\centerline{\includegraphics[width=\hsize,clip]{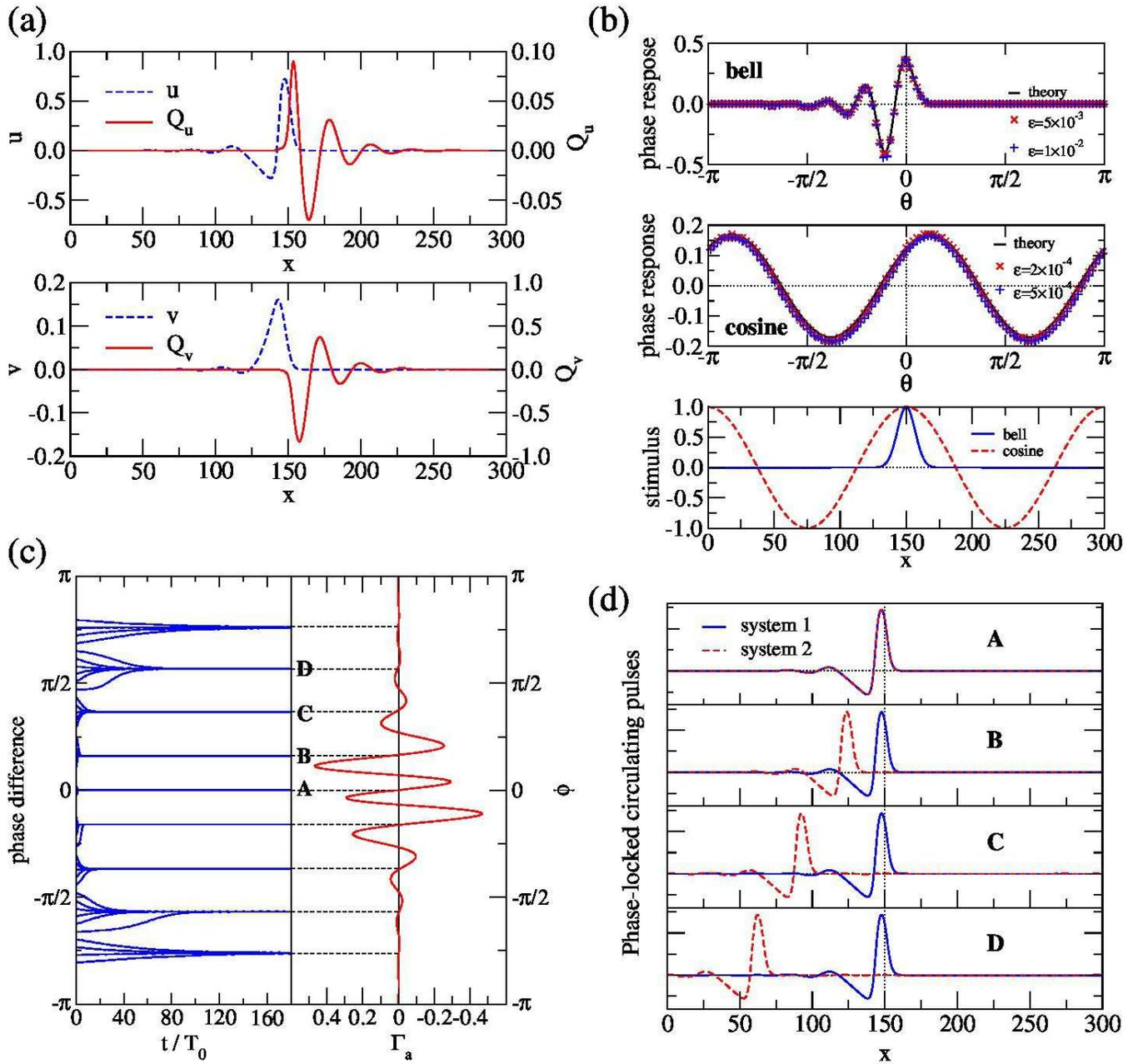}}
\caption{Circulating pulses.  The system is a 1D ring of length $L=300$ with periodic boundary conditions.  The system parameters are $\alpha = 0$, $\tau^{-1} = 0.018$, $\gamma = 1$, $\kappa = 1$, and $\delta = 0.02$.  With these values, the pulse exhibits a wavy tail~\cite{hastings}.  The oscillation period (time needed for the pulse to go around the ring) is $T \approx 1125$.
(a) Snapshots of the stable circulating pulse with a wavy tail, ${\bf X}_{0}(x ; \theta = 0) = (u(x), v(x))$, and the corresponding phase sensitivity function, ${\bf Q}(x ; \theta=0) = (Q_{u}(x), Q_{v}(x))$.
(b) Phase response curves $R(\theta)$ of the circulating pulse normalized by the stimulus intensity $\varepsilon$.  Either bell-shaped [$s(x) = \varepsilon \exp\{-(x-150)^{2} / 90\}$] or cosine [$s(x) = \varepsilon \cos (4 \pi x / 300)$] perturbation is given to the activator ($u$) component. 
(c) Evolution of phase differences between two systems coupled through the $u$ component with the coupling intensity matrix ${\rm K} = \mbox{diag}(0.001, 0)$ and the anti-symmetric part of the phase-coupling function $\Gamma_{a}(\phi)$ (rescaled by the coupling intensity $0.001$).  The two pulses show multimodal phase synchronization.
(d) Snapshots of phase-locked pulses with four different stable phase phase differences.  Graphs (A--D) correspond to the stable phase differences shown in (c).
}
\label{fig:A}
\end{figure}

We consider weakly perturbed RD systems exhibiting stable rhythmic dynamics, described by
\begin{align}
\frac{\partial}{\partial t} {\bf X}({\bf r}, t) = {\bf F}({\bf X}, {\bf r}) + {\rm D} \nabla^{2} {\bf X} + {\bf p}({\bf r}, t).
\label{eq:perturbedRD}
\end{align}
Here, the vector field ${\bf X}({\bf r}, t)$ represents the state (e.g., concentrations of chemical species) of the RD medium at point ${\bf r}$ at time $t$, ${\bf F}({\bf X},{\bf r})$ represents the local reaction dynamics at ${\bf r}$, ${\rm D} \nabla^{2} {\bf X}$ represents the diffusion of ${\bf X}$ over the medium with a matrix ${\rm D}$ of diffusion constants,
and ${\bf p}({\bf r}, t)$ represents weak spatiotemporal perturbations.
Explicit dependence of ${\bf F}$ on ${\bf r}$, such as medium heterogeneity, may exist.
We assume that the RD system~(\ref{eq:perturbedRD}) without perturbation (${\bf p}={\bf 0}$) exhibits a stable rhythmic dynamics, i.e., it possesses a stable limit-cycle solution $\chi:{\bf X}_{0}({\bf r}, t) = {\bf X}_{0}({\bf r}, t+T)$ of period $T = 2 \pi / \omega$, where $\omega$ denotes frequency, and that this solution persists and deforms only slightly even if the system is weakly perturbed (${\bf p} \neq {\bf 0}$).
Such a limit cycle includes the circulating pulses on a ring, oscillating spots, target waves,
and rotating spirals that we will analyze in Section III  (see Figs.~\ref{fig:A}, \ref{fig:B}, \ref{fig:C}, and \ref{fig:D}).

The purpose of the phase reduction theory is to derive a simple closed equation for the phase $\theta$ approximately describing limit-cycle oscillations of Eq.~(\ref{eq:perturbedRD}) under weak perturbation (${\bf p} \neq {\bf 0}$).
As in the ODE case (see Appendix A), we first introduce a phase $\theta = \omega t \ (\mbox{mod}\ 2\pi)$ to a system state ${\bf X}_{0}({\bf r}, t)$ on the limit cycle $\chi$ so that $\dot{\theta}(t) = \omega$ constantly holds, and denote the system state as ${\bf X}_{0}({\bf r} ; \theta)$ using the phase $\theta$.
To perform phase reduction, we also need to assign a phase to a system state ${\bf X}({\bf r},t)$ that is not on the limit cycle $\chi$ but eventually converges to $\chi$, because the system state can deviate from $\chi$ due to perturbations.  Specifically, we need a functional $\theta = \Theta\{{\bf X}({\bf r},t)\}$ that maps ${\bf X}({\bf r},t)$ in the basin of $\chi$ to a scalar phase $\theta$ such that $\dot{\theta}(t) = \omega$ constantly holds.
This leads to the notion of {\em isochrons}~\cite{winfree80,kuramoto,hoppensteadt97,brown04,ermentrout10,winfree67,guckenheimer}, i.e., equal-phase contours of the system state around $\chi$.
The notion of the isochrons is at the core of the conventional phase reduction theory for ODEs and should be generalized to RD systems.
It is, however, generally impossible to obtain such a functional explicitly.

To proceed, we use the assumption that the perturbation is weak and focus on the vicinity of $\chi$.  We make an ansatz that the phase $\Theta\{ {\bf X}({\bf r}) \}$ of a system state ${\bf X}({\bf r})$ near ${\bf X}_{0}({\bf r} ; \theta)$ can be linearly approximated, using a certain function  ${\bf Q}({\bf r};\theta)$, as
\begin{align}
	\Theta\{ {\bf X}({\bf r}) \} = \theta + \left[ {\bf Q}({\bf r} ; \theta),\ {\bf X}({\bf r}) - {\bf X}_{0}({\bf r}; \theta) \right]
\label{eq:approxPHS}
\end{align}
around $\chi$, where $\left[ {\bf A}({\bf r}), {\bf B}({\bf r}) \right] = \int {\bf A}({\bf r}) \cdot {\bf B}({\bf r}) d{\bf r}$ is the inner product between two functions.
For a system state ${\bf X}({\bf r}) = {\bf X}_{0}({\bf r} ; \theta)$ on $\chi$ with phase $\theta$, an identity $\Theta\{ {\bf X}_{0}({\bf r} ; \theta) \} = \theta$ should hold by the above definition of the phase.
Moreover, for any system state ${\bf X}({\bf r},t)$ near $\chi$ evolving under Eq.~(\ref{eq:perturbedRD}) with ${\bf p}={\bf 0}$, we require that $\theta(t) = \Theta\{ {\bf X}({\bf r}, t) \}$ satisfies $\dot{\theta}(t) = \omega$ constantly within linear approximation.

As we will derive in Appendix B, if ${\bf Q}({\bf r};\theta)$ is a periodic solution to a generalized {\em adjoint equation}
\begin{align}
\omega \frac{\partial}{\partial \theta} {\bf Q}({\bf r} ; \theta) = - {\rm J}( \theta )^{\dagger} {\bf Q}({\bf r} ; \theta) - {\rm D}^{\dagger} \nabla^{2} {\bf Q}({\bf r} ; \theta)
\label{eq:adjointRD}
\end{align}
with a normalization condition 
\begin{align}
\left[ {\bf Q}({\bf r};\theta), \frac{ \partial {\bf X}_{0}({\bf r} ; \theta) }{ \partial \theta } \right] = 1,
\label{eq:normalization}
\end{align}
the functional $\Theta\{{\bf X}({\bf r})\}$ assumed in Eq.~(\ref{eq:approxPHS}) satisfies the above requirements for the phase, and that such ${\bf Q}({\bf r};\theta)$ plays the role of the {\em phase sensitivity function}~\cite{winfree80,kuramoto,hoppensteadt97,ermentrout10,brown04} for the RD system.
Here, ${\rm J}(\theta) = {\rm J}( {\bf X}_{0}({\bf r} ; \theta) )$ is a Jacobi matrix of ${\bf F}$ estimated at ${\bf X} = {\bf X}_{0}({\bf r} ; \theta)$ on $\chi$.

Namely, we can show that the phase $\theta(t) = \Theta\{ {\bf X}({\bf r}, t) \}$ of the {\em infinite-dimensional} RD system~(\ref{eq:perturbedRD}) approximately obeys a simple {\em one-dimensional} phase equation
\begin{align}
\dot{\theta}(t)
&= \omega + \left[ {\bf Q}({\bf r} ; \theta),\ {\bf p}({\bf r}, t) \right],
\label{eq:phaseEQ}
\end{align}
which is correct up to the lowest order of the perturbation ${\bf p}({\bf r}, t)$.
Thus, once ${\bf Q}({\bf r} ; \theta)$ is obtained from Eqs.~(\ref{eq:adjointRD})~and~(\ref{eq:normalization}), rhythmic dynamics of the RD system subjected to weak spatiotemporal perturbations, Eq.~(\ref{eq:perturbedRD}), can easily be analyzed using Eq.~(\ref{eq:phaseEQ}).
This is the main result of the present study.

The phase equation~(\ref{eq:phaseEQ}) also shows that, if a system state ${\bf X}_{0}({\bf r} ; \theta)$ with phase $\theta$ on $\chi$ is instantaneously perturbed by a weak spatial stimulus ${\bf s}({\bf r})$, the response of the system phase after relaxation, namely, the {\em phase response curve} (PRC)~\cite{winfree80,ermentrout10}, is given by
\begin{align}
R(\theta) = [ {\bf Q}({\bf r} ; \theta),\ {\bf s}({\bf r}) ]
\end{align}
within linear approximation.
This PRC $R(\theta)$ can be directly measured in numerical simulations by applying impulsive perturbations to the RD system as we will illustrate in Section~III.

\subsection{Mutual synchronization of spatiotemporal rhythms}

As a simple example of the phase reduction approach, let us consider synchronization of a pair of weakly coupled RD systems exhibiting rhythmic dynamics,
\begin{align}
\frac{\partial}{\partial t} {\bf X}_{1}({\bf r}, t) &= {\bf F}({\bf X}_{1}, {\bf r}) + {\rm D} \nabla^{2} {\bf X}_{1} + {\bf G}\{ {\bf X}_{1}, {\bf X}_{2} \}, \cr
\frac{\partial}{\partial t} {\bf X}_{2}({\bf r}, t) &= {\bf F}({\bf X}_{2}, {\bf r}) + {\rm D} \nabla^{2} {\bf X}_{2} + {\bf G}\{ {\bf X}_{2}, {\bf X}_{1} \},
\label{eq:coupledRD}
\end{align}
where ${\bf X}_{1,2}$ represent the system states.
We assume local and linear mutual coupling, ${\bf G}\{ {\bf X}, {\bf Y} \} = {\rm K} ( {\bf Y}({\bf r}, t) - {\bf X}({\bf r}, t) )$, with a diagonal matrix ${\rm K}$ representing the intensity of the weak mutual coupling.
Experimental systems like Eq.~(\ref{eq:coupledRD}) have been realized by coupling a pair of photosensitive Belousov-Zhabotinsky chemical reactions via video cameras and projectors~\cite{hildebrand01}, and by coupling a pair of electrochemical oscillators via electrodes~\cite{fukushima}.

Denoting the phase variables of the two systems as $\theta_{1,2}$ and considering the coupling term ${\bf G}$ as weak perturbations, we can approximate Eq.~(\ref{eq:coupledRD}) by a pair of coupled phase equations,
\begin{align}
\dot{\theta}_{1}(t) &= \omega + [ {\bf Q}({\bf r} ; \theta_{1}), {\bf G}\{ {\bf X}_{0}({\bf r} ; \theta_{1}), {\bf X}_{0}({\bf r} ; \theta_{2})\} ],
\cr
\dot{\theta}_{2}(t) &= \omega + [ {\bf Q}({\bf r} ; \theta_{2}), {\bf G}\{ {\bf X}_{0}({\bf r} ; \theta_{2}), {\bf X}_{0}({\bf r} ; \theta_{1})\} ],
\label{eq:coupledPHS}
\end{align}
where ${\bf X}_{1,2}$ in ${\bf G}$ are approximated by ${\bf X}_{0}({\bf r} ; \theta_{1,2})$ as the lowest-order approximation~\cite{winfree80,kuramoto}.
Note that the {\em two infinite-dimensional} RD systems are reduced to just {\em two one-dimensional} phase equations.

The coupled phase equations~(\ref{eq:coupledPHS}) can be analyzed in the same way as those for ordinary limit cycles~\cite{winfree80,kuramoto,hoppensteadt97,brown04,ermentrout10}.  Since the coupling term ${\bf G}$ is small, we can apply the averaging method to Eqs.~(\ref{eq:coupledPHS}), which yields
\begin{align}
\dot{\theta}_{1}(t) = \omega + \Gamma(\theta_{1} - \theta_{2}),
\quad
\dot{\theta}_{2}(t) = \omega + \Gamma(\theta_{2} - \theta_{1}),
\label{eq:13}
\end{align}
where the phase-coupling function $\Gamma$ is given by
\begin{align}
\Gamma(\phi) = \frac{1}{2\pi} \int_{0}^{2\pi} \big[ {\bf Q}({\bf r} ; \theta + \phi), {\bf G}\{{\bf X}_{0}({\bf r} ; \theta + \phi), {\bf X}_{0}({\bf r} ; \theta) \} \big] \ d\theta.
\end{align}
By subtraction, the phase difference $\phi = \theta_{1} - \theta_{2}$ obeys
\begin{align}
\dot{\phi}(t) = \Gamma(\phi) - \Gamma(-\phi) = \Gamma_{a}(\phi),
\label{eq:phaseDIF}
\end{align}
where $\Gamma_{a}(\phi)$ is a $2\pi$-periodic, anti-symmetric function. 

Thus, the phase difference $\phi = \theta_{1} - \theta_{2}$ between the two RD systems approximately obeys a quite simple one-dimensional equation.
By examining the zeros of $\Gamma_{a}(\phi)$ and their stability, we can predict the stable phase differences at which phase synchronization occurs between the two limit-cycle solutions of the coupled RD systems under the phase-reduction approximation.
Since we are considering symmetrically coupled identical RD systems, $\Gamma_{a}(\phi)$  always vanishes at $\phi = 0$ and at $\phi = \pm \pi$, so that the existence of in-phase ($\phi=0$) and anti-phase ($\phi=\pm \pi$) synchronized states is assured.  Their stability is determined by the slope of $\Gamma_{a}(\phi)$.

\section{Examples}

In this section, we illustrate the phase reduction theory for RD systems
by numerical simulations.
We analyze phase response properties and synchronization dynamics of 
circulating pulses on a ring, oscillating spots, target waves, and rotating spirals
of the FitzHugh-Nagumo model of excitable media (Figs.~\ref{fig:A}-\ref{fig:D}).
Among these rhythmic patterns, the circulating pulses and rotating spirals are rigid and spatially symmetric, so that they may in principle be analyzed using the conventional methods~\cite{kuramoto,ei,maneville,cross,mori,ohta,ermentrout,lober,sandstede,biktashev}.
In contrast, the oscillating spots and target waves are not rigid and lack translational or rotational symmetry; therefore, they cannot be treated by the conventional methods that rely on such assumptions.
In any case, the phase reduction can provide a simple, unified approach to the synchronization properties of spatiotemporal rhythms.
As we will see, complex spatiotemporal profiles of the rhythmic patterns can lead to interesting synchronization dynamics.

\subsection{The FitzHugh-Nagumo model}

\begin{figure}[htbp]
\centerline{\includegraphics[width=\hsize,clip]{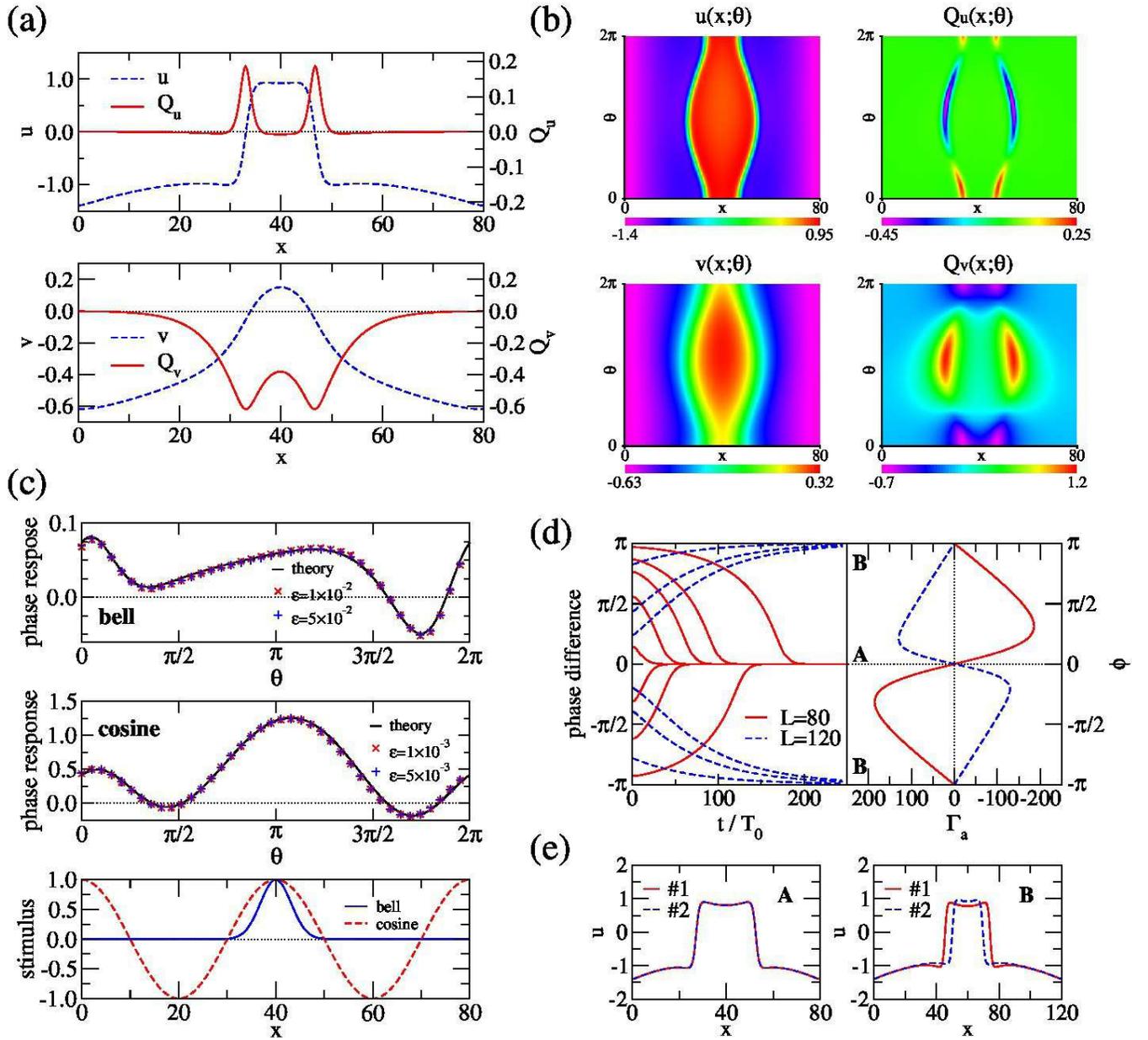}}
\caption{Oscillating spots.  The system is a 1D interval of length $L=80$ or $L=120$ with no-flux boundary conditions.  The parameter $\alpha$ is space-dependent, i.e., $\alpha(x) = \alpha_{0} + ( \alpha_{1} - \alpha_{0} ) (2x/L - 1)^{2}$ with $\alpha_{0} = -1.1$ and $\alpha_{1} = -1.6$, so that $\alpha$ is the largest at the center ($x=L/2$) and the smallest at the boundaries ($x=0,L$).
Other parameter values are $\tau^{-1} = 0.03$, $\gamma = 2.0$, $\kappa = 1$, $\delta = 2.5$.  With these conditions, an oscillating spot constrained at the center can be generated.  The oscillation period is $T=194.8$ ($L=80$) or $T=204.0$ ($L=120$).
(a) Snapshots of the oscillating spot solution ${\bf X}_{0}(x) = (u(x), v(x))$ and the corresponding phase sensitivity function ${\bf Q}(x) = (Q_{u}(x), Q_{v}(x))$ for $\theta=0$.
(b) Evolution of ${\bf X}_{0}(x;\theta)$ and ${\bf Q}(x;\theta)$ during $0 \leq \theta < 2\pi$.
(c) Phase response curves $R(\theta)$ of the oscillating spot normalized by the stimulus intensity $\varepsilon$.  Perturbation $s(x)$ is either bell-shaped [$s(x) = \varepsilon \exp\{-4 (x-L/2)^{2} / L\}$] or sinusoidal [$s(x) = \varepsilon \cos (4 \pi x / L)$] and is given to the activator ($u$) component.  Results obtained by direct numerical simulations are compared with the theory, $R(\theta) = [ {\bf Q}(x ; \theta) , {\bf s}(x) ]$, where the inner product is taken over the 1D interval ($0 \leq x \leq L$).  
(d) Evolution of phase differences between two systems coupled through the $u$ component with the intensity matrix ${\rm K} = \mbox{diag}(10^{-4}, 0)$, showing in-phase synchronization for $L=80$ (A) and anti-phase synchronization for $L=120$ (B).  The anti-symmetric part of the phase-coupling function $\Gamma_{a}(\phi)$ (rescaled by the coupling intensity $10^{-4}$) is shown for comparison.
(e) Snapshots of activator patterns $u(x)$ of both systems in the in-phase (A) and anti-phase (B) synchronized states.
}
\label{fig:B}
\end{figure}

The FitzHugh-Nagumo (FHN) reaction-diffusion model is a classical model of neural spike transmission, whose dynamics is described by
\begin{align}
{\bf X} = \left( \begin{array}{c} u \\ v \end{array} \right),
\quad
{\bf F} = \left( \begin{array}{c} u ( u - \alpha ) ( 1 - u ) - v \\ \tau^{-1} (u - \gamma v) \end{array} \right),
\quad
{\rm D} = \left( \begin{array}{cc} \kappa & 0 \\ 0 & \delta \end{array} \right),
\end{align}
where $u = u({\bf r}, t)$ and $v = v({\bf r}, t)$ are activator and inhibitor variables, respectively.
By appropriately choosing the parameters $\alpha$, $\tau$, $\gamma$, and the diffusion constants $\kappa$ and $\delta$, the FHN model can exhibit various types of rhythmic spatiotemporal dynamics~\cite{yanagita,nomura,hastings,hagberg}, such as the
circulating pulses on a ring (Fig.~\ref{fig:A}), oscillating spots (Fig.~\ref{fig:B}), target waves (Fig.~\ref{fig:C}), and rotating spirals (Fig.~\ref{fig:D}).

In numerical simulations, the size of the system is $L=80-600$ for 1D cases and discretized using $\Delta x = 0.5 - 1.0$ spatial grids.  For 2D cases, the system size is $L_{x} \times L_{y} = 80 \times 80 - 120 \times 120$, and discretized with $\Delta x = \Delta y = 0.5 -1.0$ spatial grids.
The explicit Euler method with a time step $\Delta t = 0.01 - 0.05$ is used for numerical simulations of the RD system.

To numerically obtain the phase sensitivity function ${\bf Q}({\bf r};\theta)$, the adjoint equation~(\ref{eq:adjointRD}) is integrated backward in time~\cite{ermentrout10}.  
Namely, one period of the limit-cycle oscillation is recorded by integrating the original RD system forward with sufficiently small time grids; then, the adjoint equation is spatially discretized and numerically integrated backward using the recorded time sequence of the limit cycle, with occasional normalization of the solution so that Eq.~(\ref{eq:normalization}) is satisfied.
Owing to the assumed stability of the limit-cycle solution, all modes other than the zero mode corresponding to temporal translational invariance eventually decay (the Floquet theorem), and the resultant solution gives the phase sensitivity function.

\subsection{Circulating pulses}

Our first example is a circulating-pulse solution of the FHN model with a wavy tail on a 1D ring of length $L$~\footnote{Preliminary result on this system was partially presented in our conference proceedings (not refereed)~\cite{nakao} without detailed derivation of the phase-reduction theory.}.
Since the pattern is rigid and the system is translationally symmetric, the phase $\theta$ can simply be identified as the pulse location in this case.

Figure~\ref{fig:A}(a) shows snapshots of the limit-cycle solution ${\bf X}_{0}(x ; \theta)$ and the corresponding phase sensitivity function ${\bf Q}(x ; \theta)$ for $\theta = 0$, both propagating to the right.
Results for other values of $\theta$ can simply be obtained by translating Fig.~\ref{fig:A}(a) in the $x$ direction.
It is observed that ${\bf Q}(x ; \theta)$ is {\em localized} near the pulse, indicating that perturbations given only in this region can affect the phase of the pulse.
It is also seen that ${\bf Q}(x ; \theta)$ has a {\em wavy front}, reflecting the wavy tail of the pulse.
This counterintuitive result can be explained as follows.
The system exhibits localized damped oscillations when it is perturbed at some spatial point.  If the pulse propagates into such a region, the pulse location (i.e., the system phase) is either advanced or retarded depending on the timing of the collision, yielding the wavy front of ${\bf Q}(x ; \theta)$.
Figure~\ref{fig:A}(b) compares the PRCs $R(\theta)$ of the system to the weak spatial stimulus ${\bf s}(x)$ obtained by direct numerical simulations (DNS) with the theoretical results, $R(\theta) = [{\bf Q}(x ; \theta), {\bf s}(x)]$, where ${\bf Q}(x ; \theta)$ is obtained from the adjoint equation.
The stimulus is either a bell shape localized at the center or a cosine curve, and it is given only to the activator component for the sake of simplicity.
When the intensity $\varepsilon$ of the stimulus is sufficiently small, good agreement is obtained.

Figure~\ref{fig:A}(c) shows the synchronization dynamics of the two RD systems, i.e., the evolution of the phase difference $\phi = \theta_{1} - \theta_{2}$ from various initial conditions obtained by the DNS, and compares them with the theoretical function $\Gamma_{a}(\phi)$.
Reflecting the wavy shapes of ${\bf X}_{0}$ and ${\bf Q}$, $\Gamma_{a}(\phi)$ is also wavy with many zeros, which implies the coexistence of multiple stable phase-locking points for the two-coupled circulating pulses.
This is confirmed by DNS, which shows that the final phase differences are in good agreement with the zero-crossing points of $\Gamma_{a}(\phi)$ with negative $\Gamma_{a}'(\phi)$.
Figure~\ref{fig:A}(d) shows several pairs of stably phase-locked pulses obtained by evolving the system from four different initial conditions.  The two pulses synchronize where their wavy tails match, yielding multiple stable phase differences as predicted by the phase-reduction analysis.
Similar multi-modal phase locking is also observed in complex oscillations of delay-differential systems~\cite{kotani}.

\subsection{Oscillating spots}

Our second example is an oscillating spot solution of the 1D FHN system of length $L$ with no-flux boundaries~\cite{hagberg}.
To pin the spot at the center, the parameter $\alpha$ of the model is assumed to be spatially heterogeneous, namely, the excitability of the system is the largest at the center and the smallest at the boundaries.  Note that the pattern is not rigid and the system lacks spatial symmetry.

Figure~\ref{fig:B}(a) shows snapshots of the limit-cycle solution ${\bf X}_{0}(x ; \theta)$ and the phase sensitivity function ${\bf Q}(x ; \theta)$ for $\theta=0$.  
The activator component of ${\bf Q}(x ; \theta)$ is sharply localized at both fronts of the spot; namely, the phase $\theta$ of the system is sensitive only to perturbations near the fronts.
Figure~\ref{fig:B}(b) shows ${\bf X}_{0}(x ; \theta)$ and corresponding ${\bf Q}(x ; \theta)$ for one oscillation period ($0 \leq \theta < 2\pi$).
Perturbations given to the pulse fronts result in an advance or a delay in phase, depending on the timing, i.e., whether the spot is expanding or shrinking.
The inhibitor component of ${\bf Q}(x ; \theta)$ also reflects the oscillation of the spot.
Figure~\ref{fig:B}(c) shows the PRCs $R(\theta)$ to the weak stimulus ${\bf s}(x)$, which is either a bell shape or a cosine curve and is only applied to the activator.
There is good agreement between the results of numerical simulations and theory.

The synchronization properties of a pair of oscillating spots coupled through the activator component are shown in Fig.~\ref{fig:B}(d), where the function $\Gamma_{a}(\phi)$ and evolution of the phase difference $\phi$ are plotted.
For comparison, two different system sizes, $L=80$ and $L=120$, are used.
Since the parameter $\alpha$ is spatially heterogeneous, the shape and oscillation period of the spot vary with $L$.
When $L=80$, in-phase synchronization ($\phi = 0$) is linearly stable because
$\Gamma_{a}'(\phi=0) < 0$.
In contrast, when $L=120$, in-phase synchronization is unstable and anti-phase synchronization ($\phi = \pm \pi$) becomes stable.
This prediction is confirmed by numerical simulations with various initial phase differences.
Typical snapshots of the synchronized patterns are shown in Fig.~\ref{fig:B}(e).

\begin{figure}[htbp]
\centerline{\includegraphics[width=\hsize,clip]{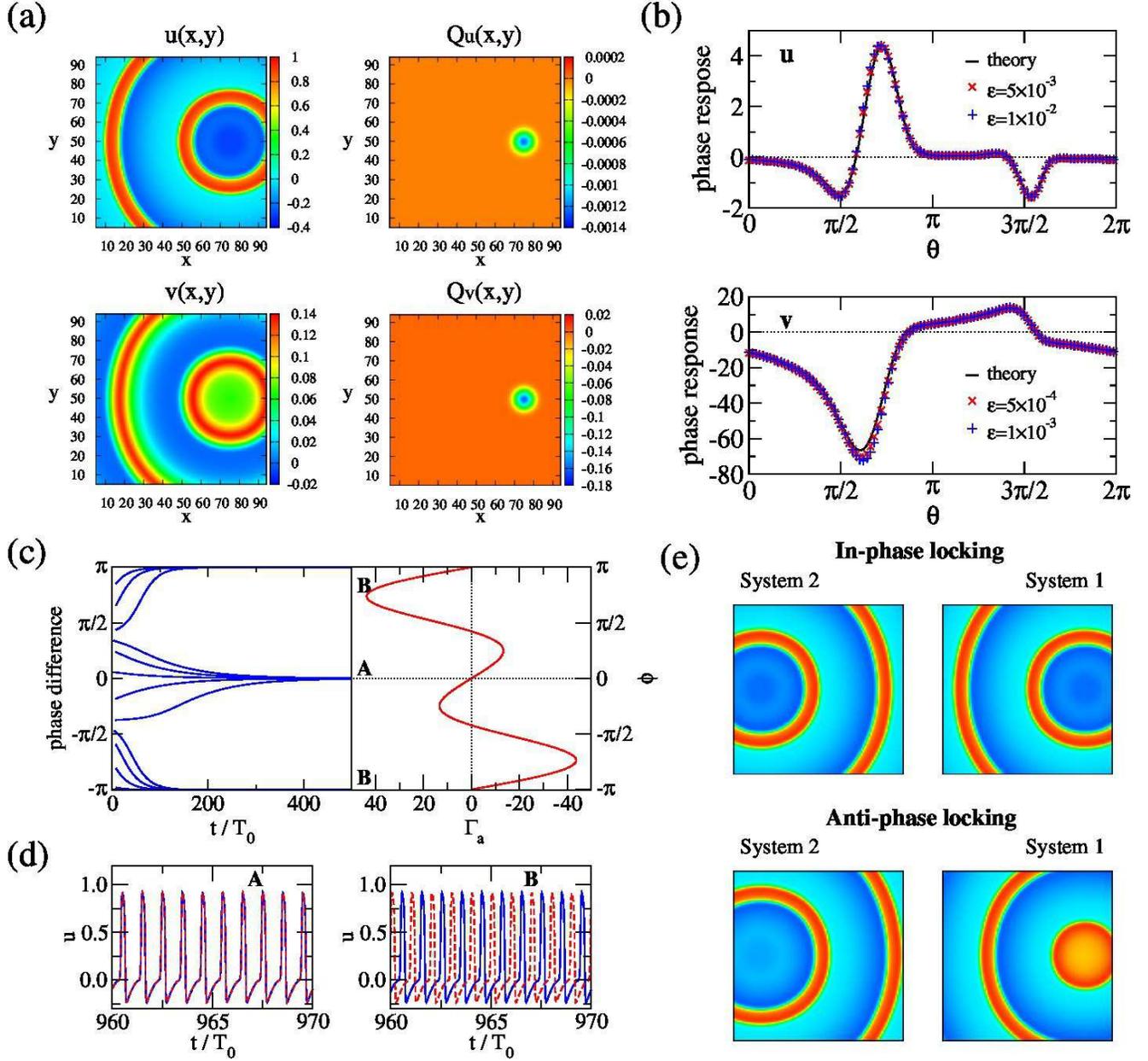}}
\caption{Target waves.  The system is a 2D square of side $L=100$ with no-flux boundary conditions.
To create a pacemaker region, the parameter $\alpha$ is assumed to possess localized circular heterogeneity, i.e., $\alpha(x,y) = \alpha_{0} + (\alpha_{1} - \alpha_{0}) \exp ( - r^{4} / r_{0}^{4} )$, where $r = [ (x - x_{0})^{2} + (y - y_{0})^{2} ]^{1/2}$ is the distance from the pacemaker center at $(x_{0}, y_{0})$ and $r_{0}$ is the radius of the pacemaker region, so that $\alpha(x, y) \to \alpha_{1}$ as $r \to 0$ and $\alpha(x,y) \to \alpha_{0}$ as $r \to \infty$.  
The parameters are $\alpha_{0} = 0.1$, $\alpha_{1} = -0.1$, $r_{0}=10$, and $(x_{0}, y_{0}) = (80, 50)$.  With these values, the system is self-oscillatory near the pacemaker center, and is excitable otherwise. Other parameters are $\tau^{-1} = 0.005$, $\gamma = 2.5$, $\kappa = 0.15$, and $\delta = 0$.  The temporal oscillation period is $T=205.4$.
(a) Target wave solution ${\bf X}_{0}(x,y) = (u(x,y), v(x,y))$ and the corresponding phase sensitivity function ${\bf Q}(x,y) = (Q_{u}(x,y), Q_{v}(x,y))$ at $\theta=0$.
(b) Phase response curves $R(\theta)$ of the target wave normalized by the stimulus intensity $\varepsilon$.  Sinusoidal perturbation $s(x,y) = \varepsilon \cos (4 \pi x / 100) \cos ( 4 \pi y / 100) $ is given either to activator ($u$) or inhibitor ($v$) component.  Results obtained by direct numerical simulations are compared with the theory, $R(\theta) = [ {\bf Q}(x, y ; \theta) , {\bf s}(x,y) ]$, where the inner product is now taken over the 2D square ($0 \leq x,y \leq L$).  
(c) Evolution of phase differences between two systems coupled through the $u$ component with the intensity matrix ${\rm K} = \mbox{diag}(5 \times 10^{-4}, 0)$, compared with the anti-symmetric part of the phase-coupling function $\Gamma_{a}(\phi)$ (rescaled by the coupling intensity $5 \times 10^{-4}$). Both in-phase (A) and anti-phase (B) synchronization can occur depending on initial conditions.  
(d) Evolution of the activator $u$ measured at the center $[(x,y)=(L/2, L/2)]$ of systems 1 and 2 in the in-phase (A) and anti-phase (B) synchronized states. Solid line corresponds to system 1, and the dashed line corresponds to system 2.
(e) Snapshots of the activator patterns $u(x,y)$ of both systems in the in-phase (A) and anti-phase (B) synchronized states.
}
\label{fig:C}
\end{figure}

\begin{figure}[htbp]
\centerline{\includegraphics[width=\hsize,clip]{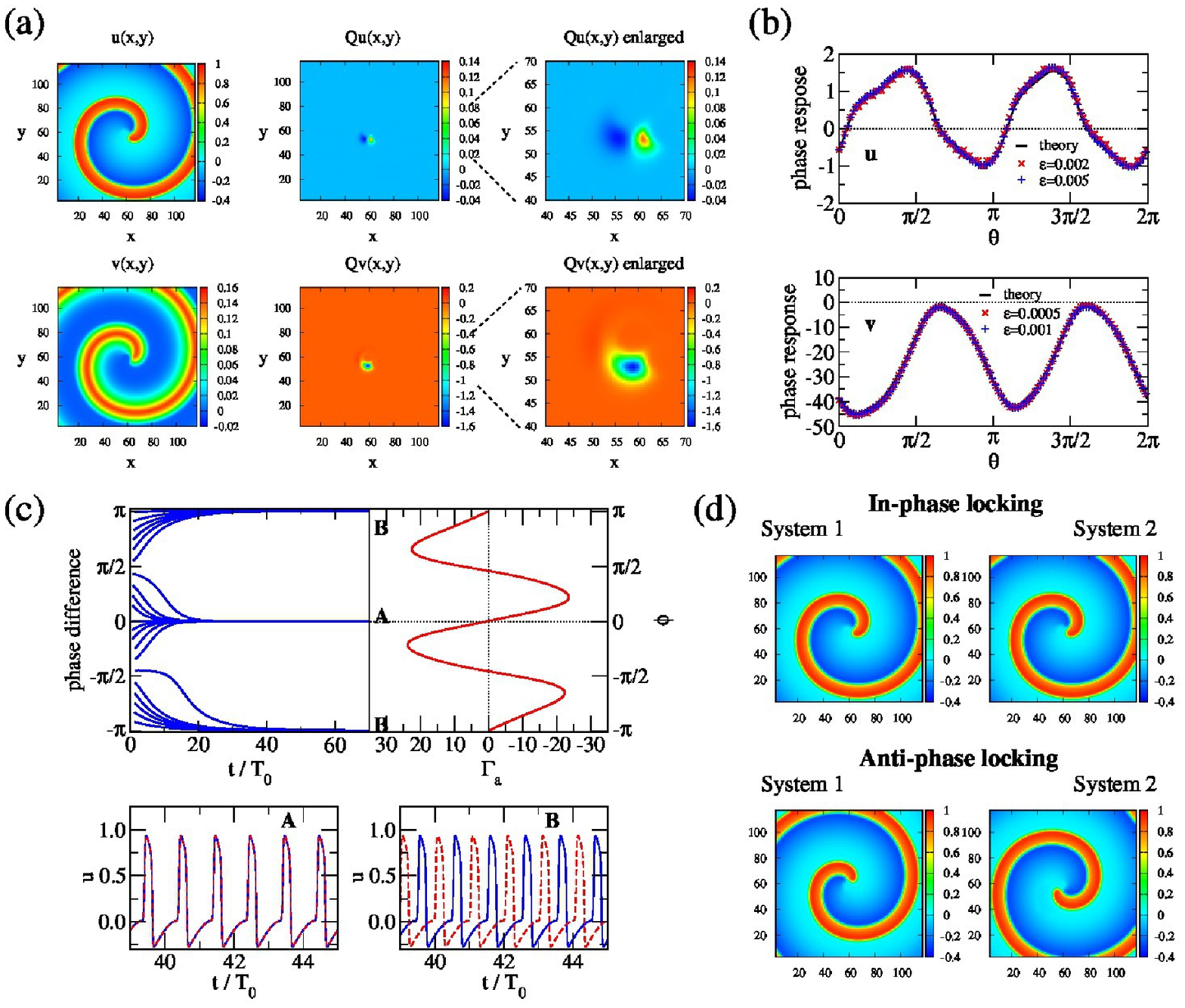}}
\caption{
(a) Rotating spirals.  The system is a 2D square of side $L=120$ with no-flux boundary conditions.  To pin the spiral at the center, a localized circular heterogeneity of radius $r_{0}=4$ is introduced to the parameter $\alpha(x,y)$ as $\alpha(x,y) = \alpha_{0} + (\alpha_{1} - \alpha_{0}) \exp( - r^{4} / r_{0}^{4} )$, where $r = [ (x - L/2)^{2} + (y - L/2)^{2} ]^{1/2}$ is a distance from center of system.  We assume $\alpha_{0} = 0.05$ and $\alpha_{1} = 0.5$, so that excitability is the highest at center.  Other parameters are fixed at $\tau^{-1} = 0.005$, $\gamma = 2.5$, $\kappa = 0.15$, and $\delta = 0$.  With these parameters, the oscillation period of the spiral is $T=217.37$.
(a) Spiral solution ${\bf X}_{0}(x,y ; \theta) = (u(x,y), v(x,y))$ and the corresponding phase sensitivity functions ${\bf Q}(x,y ; \theta) = (Q_{u}(x,y), Q_{v}(x,y))$ at $\theta=0$.
(b) Phase response curves $R(\theta)$ of the spiral normalized by the stimulus intensity $\varepsilon$.  Checkerboard-like spatial perturbation is given either to the activator ($u$) or inhibitor ($v$) component, where $s(x,y) = \varepsilon$ for $x,y>L/2$ or $x,y<L/2$, and $s(x,y)=0$ otherwise.
(c) Evolution of phase differences between two systems coupled through the $u$ component with the coupling intensity matrix ${\rm K} = \mbox{diag}(2 \times 10^{-4}, 0)$, compared with the anti-symmetric part of the phase-coupling function $\Gamma_{a}(\phi)$ (rescaled by the coupling intensity $2 \times 10^{-4}$).  Both in-phase synchronization (A) and anti-phase synchronization (B) can occur depending on initial conditions.  Solid line corresponds to system 1, and the dashed line corresponds to system 2.  (d) Snapshots of the in-phase and anti-phase synchronized states.
}
\label{fig:D}
\end{figure}

\subsection{Target waves}

As the third example, we consider a target wave solution~\cite{winfree80,kuramoto,mikhailov} of the 2D FHN model
on a square of side $L$ with no-flux boundaries.
A circular pacemaker region is created by assuming the parameter $\alpha$ to be heterogeneous.
The system is rotationally symmetric around the pacemaker region in this case, but the target pattern is not rigid; the phase should be associated with the temporal dynamics of the pattern.

Figure~\ref{fig:C}(a) shows the limit-cycle solution ${\bf X}_{0}(x, y ; \theta)$ and the corresponding phase sensitivity function ${\bf Q}(x, y ; \theta)$ for $\theta=0$.
As $\theta$ increases, ${\bf X}_{0}(x, y ; \theta)$ undergoes oscillations corresponding to the emission of concentric target waves from the pacemaker, and ${\bf Q}(x, y ; \theta)$ oscillates accordingly.
Reflecting that the pacemaker dominates overall rhythms of the system,
${\bf Q}(x, y ; \theta)$ is localized at the pacemaker.
Figure~\ref{fig:C}(b) shows the PRCs $R(\theta)$ obtained by applying weak cosine spatial stimulus ${\bf s}(x, y)$ to either the activator or inhibitor component.
The numerical results are in good agreement with the theory.

Figure~\ref{fig:C}(c) shows synchronization between two target waves.
Here, we consider counter-propagating target waves, i.e., one of the RD systems is inverted in the $x$ direction as shown in Fig.~\ref{fig:C}(e).
The function $\Gamma_{a}(\phi)$ has five zeros, with the in-phase ($\phi=0$) and anti-phase ($\phi=\pm \pi$) synchronized states both being stable.
Therefore, depending on initial conditions, the two target waves can exhibit both types of synchronization, as confirmed by numerical simulations.
Figure~\ref{fig:C}(d) shows time sequences of the activator at the center of the two systems corresponding to the in-phase and anti-phase synchronized states,
and Fig.~\ref{fig:C}(e) shows corresponding snapshots.
  
\subsection{Rotating spirals}

Our final example is a rotating-spiral solution~\cite{winfree80,kuramoto,rabinovich,mikhailov,hildebrand01} of the FHN model on a 2D square of side $L$ with no-flux boundaries.
Synchronization between a pair of rotating spirals was experimentally studied in \cite{hildebrand01}.
Here, to pin the core of the spiral at the center of the system, circular heterogeneity in the parameter $\alpha$ is introduced.
The spiral rigidly rotates around this pinning region without changing its shape.

Figure~\ref{fig:D}(a) shows snapshots of the spiral solution ${\bf X}_{0}(x, y ; \theta)$ and the corresponding phase sensitivity function ${\bf Q}(x, y ; \theta)$ at $\theta = 0$.  Both rotate in the clockwise direction as $\theta$ increases.
Since the system is symmetric with respect to spatial rotation around the center and the pattern is rigid, the phase $\theta$ simply corresponds to the rotation angle and the results for other values of $\theta$ can be obtained by rotating Fig.~\ref{fig:D}(a).
As in the other cases, ${\bf Q}(x,y ; \theta)$ is strongly localized near the core of the spiral, indicating that the spiral tip dominates the overall phase of the system; perturbations given only to this region can affect the overall system phase.
Figure~\ref{fig:D}(b) compares the PRCs obtained by DNS with the theory, $R(\theta) = [{\bf Q}(x, y ; \theta), {\bf s}(x, y)]$, to a checkerboard-like stimulus $s(x,y)$ applied either to the activator or inhibitor, showing good agreement.

Figure~\ref{fig:D}(c) shows the synchronization process between two spirals.
As expected from the function $\Gamma_{a}(\phi)$ with five zeros, the two spirals can exhibit either in-phase ($\phi=0$) or anti-phase ($\phi=\pm \pi$) synchronization as determined by the initial conditions. 
Typical time sequences of the activator component measured at $x=L/4, y=L/2$ in the in-phase and anti-phase synchronized states are shown in Fig.~\ref{fig:D}(d), and typical snapshots of the synchronized spirals are shown in Fig.~\ref{fig:D}(e).

\section{Conclusions}

We developed a phase-reduction theory for limit-cycle solutions of infinite-dimensional RD systems and illustrated its validity by analyzing mutual synchronization of a pair of RD systems exhibiting rhythmic dynamics.
Our theory does not assume rigidity and spatial symmetry; therefore, it is generally  applicable to a wide class of rhythmic spatiotemporal dynamics in RD systems.
The theory can readily be applied, for example, to the analysis of phase locking to periodic external stimulus and noise-induced synchronization~\cite{winfree80,kuramoto,hoppensteadt97,ermentrout10,pikovsky01,moehlis,harada,zlotnik,noisesync0,noisesync1,noisesync2} of spatiotemporal rhythms, and will be a basis for controlling and designing spatiotemporal rhythmics in various systems.

\clearpage

\appendix

\section{Phase reduction of ordinary limit-cycle oscillators}

In this section, we review the classical phase reduction theory for ordinary limit-cycle oscillators described by finite-dimensional ODEs.  See Refs.~\cite{winfree80,kuramoto,hoppensteadt97,ermentrout10,brown04} for details of the theory and applications to various synchronization phenomena in systems of coupled oscillators.

\subsection{Geometric formulation of the phase reduction theory}

We consider a limit-cycle oscillator described by the ODE
\begin{align}
\dot{\bf X}(t) = {\bf F}({\bf X}(t)),
\label{seq1}
\end{align}
where ${\bf X}(t)$ is a $d \geq 2$ dimensional vector representing the oscillator state at time $t$ and ${\bf F}$ determines its dynamics.
Suppose that Eq.~(\ref{seq1}) has a stable limit-cycle solution of period $T$, 
\begin{align}
\chi : {\bf X}_{0}(t) = {\bf X}_{0}(t+T),
\end{align}
which is denoted as $\chi$.
We introduce a phase $\theta(t) \in [0,2\pi)$ to the state ${\bf X}_{0}(t)$ on $\chi$ in such a way that $\theta(t)$ increases with a constant frequency $\omega = 2 \pi / T$ as ${\bf X}_{0}(t)$ evolves along $\chi$ under Eq.~(\ref{seq1}).
This can be performed by choosing a certain state ${\bf X}_{0}(t=0)$ on $\chi$ as the origin of the phase, i.e., $\theta = 0$, and assigning a phase value
\begin{align}
\theta = \omega t \;(\mbox{mod}\; 2\pi)
\end{align}
to the oscillator state ${\bf X}_{0}(t)$ on $\chi$ ($t\geq0$) evolving under Eq.~(\ref{seq1}) from the phase origin ${\bf X}_{0}(t=0)$.
Namely, we identify the oscillator phase with the time multiplied by the frequency.
We will denote the oscillator state on $\chi$ with the phase value $\theta$ as ${\bf X}_{0}(\theta)$ henceforth.

\begin{figure}[htbp]
\centering
\includegraphics[width=0.8\hsize,clip]{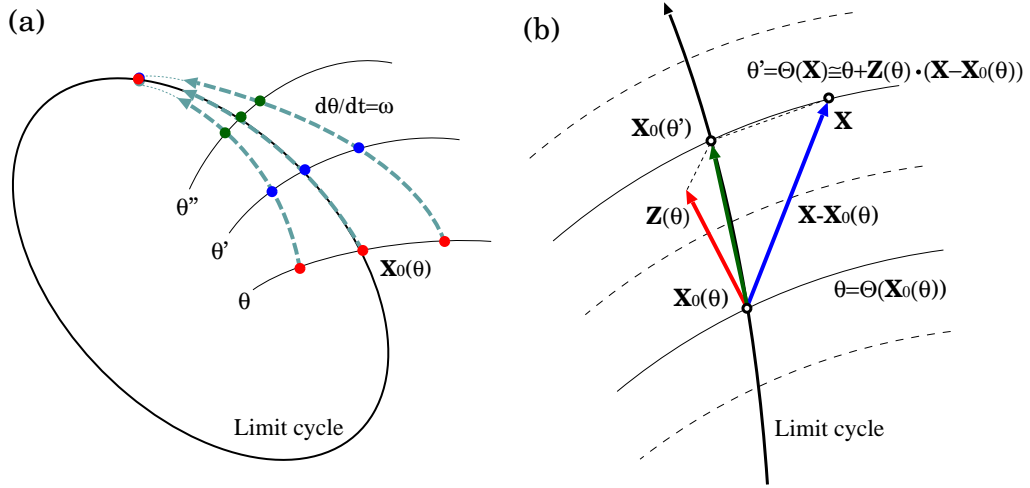}
\caption{(a) Isochrons of a limit cycle.  The same phase value is assigned to the oscillator states that asymptotically converge to the same state on the limit cycle.  (b) Linear approximation of the phase function near the limit cycle.}
\label{figS0ODE}
\end{figure}

The above definition of the phase on $\chi$ can be extended to the whole basin of $\chi$ by assigning the same phase value $\theta(t)$ to the set of oscillator states $\{ {\bf X}(t) \}$ that asymptotically approach the oscillator state ${\bf X}_{0}(\theta(t))$ on $\chi$ under Eq.~(\ref{seq1}), i.e.,
\begin{align}
\lim_{t \to +\infty} | {\bf X}(t) - {\bf X}_{0}(\theta(t)) | = 0,
\end{align}
where $| \cdots |$ represents the ordinary vector norm.
This defines a phase function
\begin{align}
\theta = \Theta({\bf X}) \in [0,2\pi)
\end{align}
that maps a given oscillator state ${\bf X}$ in the basin of $\chi$ to a scalar phase $\theta$.
It is clear that the phase $\theta(t) = \Theta({\bf X}(t))$ of the state ${\bf X}(t)$ evolving under Eq.~(\ref{seq1}) obeys a simple phase equation,
\begin{align}
\dot{\theta}(t) = \omega,
\end{align}
not only on the limit-cycle solution $\chi$ but also in the whole basin of $\chi$.
Using the chain rule for the derivatives, it can be shown that
\begin{align}
\dot{\theta}(t) = \frac{d}{dt} \Theta({\bf X}(t))
=
\left. \frac{\partial \Theta({\bf X})}{\partial {\bf X}} \right|_{{\bf X} = {\bf X}(t)} \cdot \frac{d{\bf X}(t)}{dt}
=
\left. \frac{\partial \Theta({\bf X})}{\partial {\bf X}} \right|_{{\bf X} = {\bf X}(t)} \cdot {\bf F}({\bf X}(t)) = \omega,
\label{eq:S7}
\end{align}
where $\partial \Theta({\bf X}) / \partial {\bf X}|_{{\bf X} = {\bf X}(t)}$ is the gradient of the phase function $\Theta({\bf X})$ at ${\bf X} = {\bf X}(t)$.
Thus, the phase function $\Theta({\bf X})$ should satisfy
\begin{align}
\frac{\partial \Theta({\bf X})}{\partial {\bf X}} \cdot {\bf F}({\bf X}) = \omega
\end{align}
in the basin of $\chi$.
The set of oscillator states sharing the same phase value is called the {\em isochron}~\cite{winfree67,guckenheimer} and is the fundamental concept in the analysis of limit-cycle oscillators~\cite{winfree80,kuramoto,hoppensteadt97,ermentrout10,brown04}.
The whole basin of $\chi$ is foliated by such isochrons.
See Fig.~\ref{figS0ODE}(a) for a schematic illustration of the isochrons.\\

Now we consider the case that the limit-cycle oscillator is weakly perturbed as
\begin{align}
\dot{\bf X}(t) = {\bf F}({\bf X}(t)) + {\bf p}({\bf X}(t), t),
\label{seq7}
\end{align}
where the perturbation ${\bf p}({\bf X}, t)$, generally a function of the oscillator state ${\bf X}$ and time $t$, is assumed to be sufficiently weak so that the original limit-cycle solution $\chi$ is only slightly deformed.
From Eq.~(\ref{eq:S7}), the phase $\theta(t) = \Theta({\bf X}(t))$ of the perturbed oscillator obeys
\begin{align}
\dot{\theta}(t) = \left. \frac{\partial \Theta({\bf X})}{\partial {\bf X}} \right|_{{\bf X} = {\bf X}(t)} \cdot \left\{ {\bf F}({\bf X}(t)) + {\bf p}({\bf X}, t) \right\}
=
\omega + \left. \frac{\partial \Theta({\bf X})}{\partial {\bf X}} \right|_{{\bf X} = {\bf X}(t)} \cdot {\bf p}({\bf X}(t), t).
\label{seqX1}
\end{align}
However, this is not a closed equation for $\theta(t)$ because the gradient
$\partial \Theta({\bf X}) / \partial {\bf X}|_{{\bf X} = {\bf X}(t)}$ and the perturbation ${\bf p}({\bf X}(t), t)$ still depend on ${\bf X}(t)$.
To obtain a closed equation for $\theta(t)$,
${\bf X}(t)$ in these terms are replaced by ${\bf X}_{0}(\theta(t))$ at the lowest order approximation, assuming that the perturbation ${\bf p}({\bf X}(t), t)$ is sufficiently weak so that ${\bf X}(t)$ does not significantly deviate from ${\bf X}_{0}(\theta(t))$ on $\chi$.  This yields an approximate closed phase equation for $\theta(t)$, 
\begin{align}
\dot{\theta}(t)
\simeq
\omega + \left. \frac{\partial \Theta({\bf X})}{\partial {\bf X}} \right|_{{\bf X} = {\bf X}_{0}(\theta(t))} \cdot {\bf p}({\bf X}_{0}(\theta(t)), t),
\end{align}
which is correct up to $O(|{\bf p}|)$.
Thus, by denoting ${\bf p}(\theta, t) = {\bf p}({\bf X}_{0}(\theta), t)$ and introducing a function
\begin{align}
{\bf Z}(\theta)
= \left. \frac{\partial \Theta({\bf X})}{\partial {\bf X}} \right|_{{\bf X} = {\bf X}_{0}(\theta)},
\label{seq8}
\end{align}
the $d$-dimensional ODE~(\ref{seq7}) describing a perturbed limit cycle can be reduced to a simple one-dimensional phase equation
\begin{align}
\dot{\theta}(t) = \omega + {\bf Z}(\theta(t)) \cdot {\bf p}(\theta(t), t)
\label{seq9}
\end{align}
at the lowest order in the perturbation.

The key quantity for this approximation is the {\em phase sensitivity function} ${\bf Z}(\theta)$ defined in Eq.~(\ref{seq8}), which is the gradient of the isochron estimated at ${\bf X} = {\bf X}_{0}(\theta)$ on the limit-cycle solution $\chi$.
The function ${\bf Z}(\theta)$ quantifies linear phase response property of the oscillator state ${\bf X}_{0}(\theta)$ with phase $\theta$ on $\chi$ to infinitesimal perturbations.
If ${\bf X}_{0}(\theta)$ is instantaneously perturbed by a weak stimulus ${\bf s}$, the resulting phase response is given by
\begin{align}
	R(\theta) = {\bf Z}(\theta) \cdot {\bf s}
\end{align}
under linear approximation.  This $R(\theta)$ is called the {\em phase response curve} (PRC) of the limit-cycle oscillator described by Eq.~(\ref{seq1}).  The PRC can be obtained by applying impulsive perturbations to a limit-cycle oscillator and has been measured in various experimental systems~\cite{winfree80}.

\subsection{Linear theory around the limit-cycle solution}

Though we have developed a formal geometric theory by assuming the existence of the phase function $\Theta({\bf X})$, it is generally impossible to obtain $\Theta({\bf X})$ explicitly, except for a few simple models of limit-cycle oscillators.
However, to obtain the lowest-order phase equation~(\ref{seq8}) for weak perturbations, only the phase sensitivity function ${\bf Z}(\theta)$ is actually necessary.
As we show below, the function ${\bf Z}(\theta)$ can be obtained as the $2\pi$-periodic solution to the following {\em adjoint equation}~\cite{hoppensteadt97,ermentrout10}:
\begin{align}
\omega \frac{d{\bf Z}(\theta)}{d\theta} = - {\rm J}(\theta)^{\dagger} {\bf Z}(\theta),
\label{seq12}
\end{align}
with the constraint ${\bf Z}(\theta) \cdot {\bf F}({\bf X}_{0}(\theta)) = \omega$, or equivalently,
\begin{align}
{\bf Z}(\theta) \cdot \frac{d}{d \theta} {\bf X}_{0}(\theta) = 1,
\label{seq13}
\end{align}
for $0 \leq \theta < 2\pi$, where ${\rm J}(\theta) = {\rm J}({\bf X}_{0}(\theta))$ is the Jacobi matrix of ${\bf F}({\bf X})$ at ${\bf X} = {\bf X}_{0}(\theta)$ on $\chi$ and $^{\dagger}$ indicates matrix transpose.

The adjoint equation~(\ref{seq12}) and the normalization condition~(\ref{seq13}) can be derived in several different ways.  We here use a simple argument as in~\cite{ermentrout10,brown04} with an emphasis on the linear approximation of the isochrons near the limit cycle (see Eq.~(\ref{seqY1}) below).  We use the same idea to develop a phase reduction theory for the limit-cycle solutions of reaction-diffusion systems in Appendix B.  See Fig.~\ref{figS0ODE}(b) for a schematic illustration.   

We first note that, when $| {\bf X} - {\bf X}_{0}(\theta) |$ is sufficiently small, the phase function $\Theta({\bf X})$ for the ODE can be expanded in a Taylor series around ${\bf X}_{0}(\theta)$ as
\begin{align}
\Theta( {\bf X} )
&= \Theta( {\bf X}_{0}(\theta) + {\bf X} - {\bf X}_{0}(\theta) ) \cr
&= \Theta( {\bf X}_{0}(\theta) ) + \left. \frac{\partial \Theta({\bf X})}{\partial {\bf X}} \right|_{{\bf X} = {\bf X}_{0}(\theta)} \cdot ( {\bf X} - {\bf X}_{0}(\theta) ) + O(|{\bf X} - {\bf X}_{0}(\theta)|^{2} ) \cr
&= \theta + {\bf Z}(\theta) \cdot ( {\bf X} - {\bf X}_{0}(\theta) ) + O(|{\bf X} - {\bf X}_{0}(\theta)|^{2} )
\end{align}
using the phase sensitivity function ${\bf Z}(\theta)$ in Eq.~(\ref{seq8}).
Therefore, if the oscillator state ${\bf X}$ is close to the oscillator state ${\bf X}_{0}(\theta)$ with phase $\theta$ on the limit-cycle solution $\chi$, $\Theta({\bf X})$ can be linearly approximated as
\begin{align}
\Theta({\bf X}) \simeq \theta + {\bf Z}(\theta) \cdot ( {\bf X} - {\bf X}_{0}(\theta) ).
\label{seqY1}
\end{align}

Suppose a initial state ${\bf X}_{0}(t=0) = {\bf X}(\theta = 0)$ on $\chi$, and a slightly perturbed initial state
\begin{align}
{\bf X}(t=0) = {\bf X}_{0}(\theta=0) + {\bf y}(t=0)
\end{align}
near ${\bf X}_{0}(\theta=0)$, where ${\bf y}(t=0)$ is a small perturbation given to ${\bf X}_{0}(\theta=0)$.
We evolve these two states without applying further perturbations.
Then, from Eq.~(\ref{seq1}), the linearized equation for the small perturbation ${\bf y}(t) = {\bf X}(t) - {\bf X}_{0}(\theta(t))$ is given by
\begin{align}
\frac{d}{dt} {\bf y}(t) = {\rm J}(\theta(t)) {\bf y}(t),
\end{align}
where ${\rm J}(\theta(t)) = {\rm J}({\bf X}_{0}(\theta(t)) = {\bf DF}({\bf X}_{0}(\theta(t))$ is the Jacobi matrix of ${\bf F}({\bf X})$ at ${\bf X} = {\bf X}_{0}(\theta)$ on $\chi$.
From Eq.~(\ref{seqY1}), the phase of the unperturbed state is given by $\theta(t) = \Theta({\bf X}_{0}(\theta(t))) = \omega t$, and that of the perturbed state can be expressed as
\begin{align}
\theta'(t) = \Theta({\bf X}(t)) = \Theta({\bf X}_{0}(\theta(t)) + {\bf y}(t)) \simeq \theta(t) + {\bf Z}(\theta(t)) \cdot {\bf y}(t)
\end{align}
under the linear approximation.
Note here that $\theta'(t)$ should also increase with a constant frequency $\omega$, i.e., $d\theta'(t) / dt = \omega$ within the linear approximation, because no perturbation is given after $t=0$.
Thus, the following equation should hold for ${\bf y}(t)$ evolving from arbitrary ${\bf y}(t=0)$:
\begin{align}
0
&= \frac{d}{dt} \left\{ \theta'(t) - \theta(t) \right\} = \frac{d}{dt} \left\{ {\bf Z}(\theta(t)) \cdot {\bf y}(t) \right\}
=
\frac{d {\bf Z}(\theta(t)) }{dt} \cdot {\bf y}(t) + {\bf Z}(\theta(t)) \cdot \frac{d{\bf y}(t) }{dt} \nonumber\\[18pt]
&=
\frac{d {\bf Z}(\theta(t)) }{dt} \cdot {\bf y}(t) + {\bf Z}(\theta(t)) \cdot  {\rm J}(\theta(t)) {\bf y}(t)
=
\left\{ \frac{d}{dt} {\bf Z}(\theta(t)) + {\rm J}(\theta(t))^{\dag} {\bf Z}(\theta(t)) \right\} \cdot {\bf y}(t).
\end{align}
Therefore, ${\bf Z}(\theta(t))$ should satisfy the following adjoint equation:
\begin{align}
\frac{d}{dt} {\bf Z}(\theta(t)) = - {\rm J}(\theta(t))^{\dag} {\bf Z}(\theta(t)),
\end{align}
which is equivalent to Eq.~(\ref{seq12}) by the relation $d / dt = \omega d / d\theta$ (note that $\theta = \omega t$).
To obtain the normalization condition Eq.~(\ref{seq13}), we differentiate the identity $\theta(t) = \Theta({\bf X}_{0}(\theta(t))) = \omega t$ by $t$, which yields
\begin{align}
\omega = \frac{d}{dt} \theta(t) = \left. \frac{\partial \Theta({\bf X})}{\partial {\bf X}} \right|_{{\bf X} = {\bf X}_{0}(\theta(t))} \cdot \frac{d}{dt} {\bf X}_{0}(\theta(t)) = {\bf Z}(\theta(t)) \cdot {\bf F}({\bf X}_{0}(\theta(t))).
\end{align}
This gives the normalization condition Eq.~(\ref{seq13}), again by the relation $d / dt = \omega d / d\theta$.

Thus, the function ${\bf Z}(\theta)$ can be obtained by solving the adjoint equation~(\ref{seq12}) under the normalization condition Eq.~(\ref{seq13}), and the phase function near $\chi$ is given by Eq.~(\ref{seqY1}) within linear approximation.
It can also be shown that ${\bf Z}(\theta)$ is the unique solution to Eq.~(\ref{seq12}) by using the Floquet theorem characterizing the linear stability of the limit cycle, since ${\bf Z}(\theta)$ is closely related to the Floquet eigenvector with the zero Floquet exponent~\cite{kuramoto,ermentrout10,brown04,hoppensteadt97}.
In actual numerical calculations, it is useful to integrate Eq.~(\ref{seq12}) backward in time  to avoid numerical overflow, with occasional normalization using Eq.~(\ref{seq13})~\cite{ermentrout10}.
Then, by virtue of the Floquet theorem, only the functional component corresponding to ${\bf Z}(\theta)$ remains numerically.

Once we obtain the frequency $\omega$ and the phase sensitivity function ${\bf Z}(\theta)$, we can write down the approximate phase equation~(\ref{seq9}) for a weakly perturbed limit-cycle oscillator described by Eq.~(\ref{seq7}).
This approximation, called the {\em phase reduction}, greatly simplifies theoretical analysis of weakly perturbed limit cycles and has been extensively used for analyzing synchronization dynamics of weakly interacting nonlinear oscillators~\cite{winfree80,kuramoto,hoppensteadt97,ermentrout10,brown04}.

\section{Phase reduction theory for reaction-diffusion systems}

In this section, we give a full derivation of the phase reduction theory for RD systems that we briefly summarized in Section~II~A.
Our aim is to derive a simple one-dimensional phase equation for rhythmic spatiotemporal patterns described as limit-cycle solutions of RD systems without recourse to spatial symmetry of the patterns.
We do not require the patterns to be rigidly translating or rotating in the RD medium without changing their spatial profiles, as typically assumed in the conventional derivation of the phase equations for RD systems.
Such rhythmic patterns can thus include oscillating spots and target waves, which vary their spatial profiles periodically.
Rigidly circulating waves or rotating spirals with spatial translational or rotational symmetry are also limit-cycle solutions of RD systems, and thus they can also be treated in the same framework as we showed in Section III.

Our strategy is to generalize the conventional phase reduction theory for ordinary limit cycles described by ODEs (see Appendix A for comparison), which assumes only temporal translational symmetry of the oscillator dynamics, to limit-cycle solutions of infinite-dimensional RD systems, thereby avoiding the assumptions on spatial symmetry.
We can develop the theory almost in parallel with the ODE case by noticing that the finite-dimensional vector ${\bf X}(t)$ is replaced by a vector field ${\bf X}({\bf r}, t)$, and correspondingly the ordinary dot product of two vectors is replaced by the inner product of two vector fields.

\subsection{Geometric formulation of the phase reduction theory}

We consider a RD equation of the form
\begin{align}
\frac{\partial}{\partial t} {\bf X}({\bf r}, t) = {\bf F}({\bf X}({\bf r}, t), {\bf r}) + {\rm D} \nabla^{2} {\bf X}({\bf r}, t),
\label{seq14}
\end{align}
where the $d$-dimensional vector ${\bf X}({\bf r}, t)$ represents the state of the RD medium at point ${\bf r}$ in the $n$-dimensional space at time $t$, ${\bf F}({\bf X},{\bf r})$ specifies local reaction dynamics at point ${\bf r}$, and ${\rm D} \nabla^{2} {\bf X}({\bf r}, t)$ represents diffusion of ${\bf X}$ over the medium with a constant diffusion matrix ${\rm D}$.
Explicit dependence of ${\bf F}$ on ${\bf r}$, such as heterogeneity of the medium, may exist.  Appropriate boundary conditions (e.g., periodic or no-flux) for the problem under consideration are introduced.
We assume that Eq.~(\ref{seq14}) has a stable limit-cycle solution of period $T$, 
\begin{align}
\chi : {\bf X}_{0}({\bf r}, t) = {\bf X}_{0}({\bf r}, t+T),
\end{align}
which is denoted by $\chi$.
As in the ODE case, we first define a phase $\theta(t) \in [0, 2\pi)$ of the system state ${\bf X}_{0}({\bf r}, t)$ on the limit-cycle solution $\chi$ so that $\theta(t)$ increases with a constant frequency $\omega = 2 \pi / T$ as ${\bf X}_{0}({\bf r}, t)$ evolves on $\chi$ under Eq.~(\ref{seq14}).
This is performed by identifying the phase with the time multiplied by the frequency.  Namely, we choose a certain system state ${\bf X}_{0}({\bf r}, t=0)$ on $\chi$ as the origin of the phase, $\theta = 0$, and assign a phase value
\begin{align}
\theta = \omega t \;(\mbox{mod}\; 2\pi)
\end{align}
to a state ${\bf X}_{0}({\bf r}, t)$ on $\chi$ ($t \geq 0$) evolving under Eq.~(\ref{seq14}) from the phase origin ${\bf X}_{0}({\bf r}, t=0)$.
We will denote the system state on $\chi$ with the phase value $\theta$ as ${\bf X}_{0}({\bf r} ; \theta)$ henceforth.\\

\begin{figure}[htbp]
\centering
\includegraphics[width=0.8\hsize,clip]{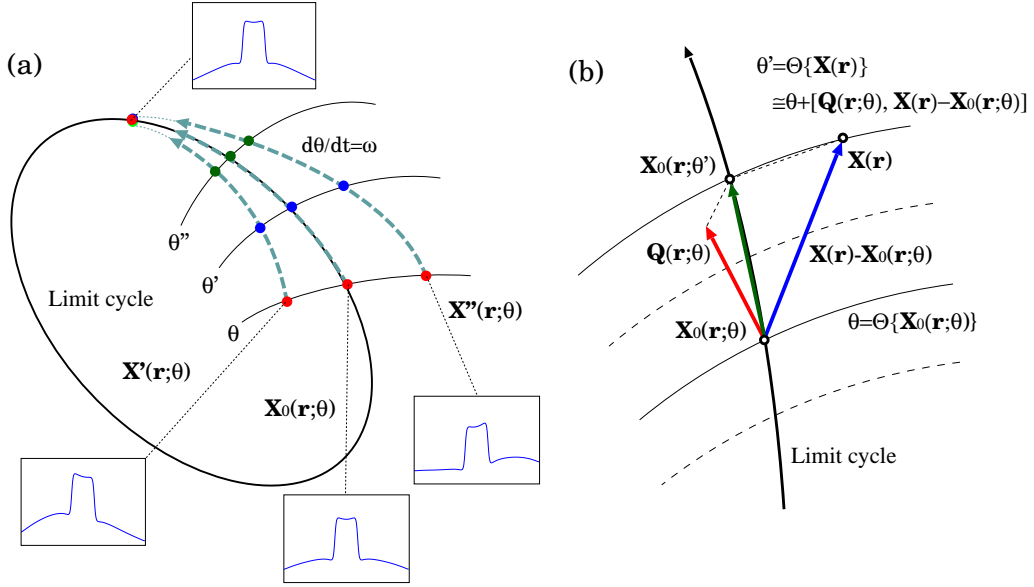}
\caption{(a) Isochrons of a limit-cycle solution of a reaction-diffusion system.  The same phase value is assigned to the system states (represented by vector fields) that asymptotically converge to the same state on the limit-cycle solution.  (b) Linear approximation of the phase near the limit-cycle solution.}
\label{figS0RD}
\end{figure}

Next, we need to extend the definition of the phase to the basin of $\chi$.
As in the ODE case, we assign the same phase value $\theta(t)$ to the set of system states $\{ {\bf X}({\bf r}, t) \}$ that eventually converge to the system state ${\bf X}_{0}({\bf r} ; \theta(t))$ on $\chi$ under Eq.~(\ref{seq14}), namely,
\begin{align}
\lim_{t \to +\infty} || {\bf X}({\bf r}, t) - {\bf X}_{0}({\bf r} ; \theta(t)) || = 0.
\end{align}
Here, $|| \cdots ||$ denotes the $L^{2}$ norm of a spatial pattern defined as $|| {\bf A}({\bf r})||^{2} = [ {\bf A}({\bf r}), {\bf A}({\bf r})]$, and the inner product between two spatial patterns ${\bf A}({\bf r})$ and ${\bf B}({\bf r})$ is defined as
\begin{align}
\left[ {\bf A}({\bf r}), {\bf B}({\bf r}) \right] = \int {\bf A}({\bf r}) \cdot {\bf B}({\bf r}) d{\bf r}.
\end{align}
The integral is taken over the considered spatial domain with appropriate boundary conditions.
This introduces a {\em phase functional}
\begin{align}
\theta = \Theta\{ {\bf X}({\bf r}) \}  \in [0,2\pi)
\end{align}
that maps a given system state ${\bf X}({\bf r})$ in the basin of $\chi$ to a scalar phase $\theta$.
Then, the phase $\theta(t) = \Theta\{ {\bf X}({\bf r}, t) \}$ of the state ${\bf X}({\bf r}, t)$ evolving under Eq.~(\ref{seq14}) will constantly obey 
\begin{align}
\dot{\theta}(t) = \omega
\end{align}
not only on the limit-cycle solution $\chi$ but also in the whole basin of $\chi$.
Using the chain rule for the functional derivatives, the above equation can be written as
\begin{align}
\dot{\theta}(t)
= \frac{d}{dt} \Theta\{ {\bf X}({\bf r}, t) \}
&= \left[ \left. \frac{\delta \Theta\{ {\bf X}({\bf r}) \} }{\delta {\bf X}({\bf r})} \right|_{{\bf X}({\bf r})= {\bf X}({\bf r}, t)}, \frac{\partial}{\partial t} {\bf X}({\bf r}, t) \right] \cr
&= \left[ \left. \frac{\delta \Theta\{ {\bf X}({\bf r}) \} }{\delta {\bf X}({\bf r})} \right|_{{\bf X}({\bf r})= {\bf X}({\bf r}, t)}, {\bf F}({\bf X}({\bf r}, t), {\bf r}) + {\rm D} \nabla^{2} {\bf X}({\bf r}, t) \right]
= \omega,
\label{seq32}
\end{align}
where $\delta \Theta\{ {\bf X}({\bf r}) \} / \delta {\bf X}({\bf r})$ is the functional derivative of $\Theta\{ {\bf X}({\bf r}) \}$ with respect to ${\bf X}({\bf r}) = {\bf X}({\bf r}, t)$.
Thus, the phase functional $\Theta\{ {\bf X}({\bf r}) \}$ should satisfy
\begin{align}
\left[ \frac{\delta \Theta\{ {\bf X}({\bf r}) \} }{\delta {\bf X}({\bf r})},\ {\bf F}({\bf X}({\bf r}), {\bf r}) + {\rm D} \nabla^{2} {\bf X}({\bf r}) \right]
= \omega
\end{align}
in the basin of $\chi$.
We call a set of system states sharing the same phase value the {\em isochron} of the RD system, generalizing the same notion for ODEs (Appendix A).  The whole basin of $\chi$ is foliated by such isochrons.
See Fig.~\ref{figS0RD}(a) for a schematic illustration.

Now we consider the case that the RD system is weakly perturbed as
\begin{align}
\frac{\partial}{\partial t} {\bf X}({\bf r}, t) = {\bf F}({\bf X}({\bf r}, t), {\bf r}) + {\rm D} \nabla^{2} {\bf X}({\bf r}, t) + {\bf p}\{ {\bf X}({\bf r}, t), {\bf r}, t \},
\label{seqP1}
\end{align}
where the perturbation ${\bf p}\{ {\bf X}({\bf r}, t), {\bf r}, t \}$ is generally a functional of the state ${\bf X}({\bf r}, t)$, location ${\bf r}$, and time $t$.
We assume that the original limit-cycle solution $\chi$ is only slightly deformed by the perturbation ${\bf p}$.
If the phase functional $\Theta\{ {\bf X}({\bf r}) \}$ is given, then from Eq.~(\ref{seq32}), the phase $\theta(t) = \Theta\{ {\bf X}({\bf r}, t) \}$ of the perturbed system obeys
\begin{align}
\dot{\theta}(t)
= \frac{d}{dt} \Theta\{ {\bf X}({\bf r}, t) \}
&= \left[ \left. \frac{\delta \Theta\{ {\bf X}({\bf r}) \} }{\delta {\bf X}({\bf r})} \right|_{{\bf X}({\bf r})= {\bf X}({\bf r}, t)}, {\bf F}({\bf X}({\bf r}, t), {\bf r}) + {\rm D} \nabla^{2} {\bf X}({\bf r}, t) + {\bf p}\{ {\bf X}({\bf r}, t), {\bf r}, t \} \right] \cr
&= \omega + \left[ \left. \frac{\delta \Theta\{ {\bf X}({\bf r}) \} }{\delta {\bf X}({\bf r})} \right|_{{\bf X}({\bf r})= {\bf X}({\bf r}, t)}, {\bf p}\{ {\bf X}({\bf r}, t), {\bf r}, t \} \right].
\end{align}
However, this is not a closed equation for $\theta(t)$ because the functional derivative of $\Theta\{ {\bf X}({\bf r}, t) \}$ and the perturbation ${\bf p}\{ {\bf X}({\bf r}, t), {\bf r}, t \}$ still depend on ${\bf X}({\bf r}, t)$.
Therefore, as in the ODE case, we approximate ${\bf X}({\bf r}, t)$ in these terms by ${\bf X}_{0}({\bf r} ; \theta)$ on $\chi$, assuming the perturbation ${\bf p}\{ {\bf X}({\bf r}, t), {\bf r}, t \}$ to be weak enough so that the system state ${\bf X}({\bf r}, t)$ deviates from ${\bf X}_{0}({\bf r} ; \theta)$ on $\chi$ only slightly.
Then, at the lowest order approximation, a closed phase equation for $\theta(t)$ can be obtained as
\begin{align}
\dot{\theta}(t)
\simeq \omega + \left[ \left. \frac{\delta \Theta\{ {\bf X}({\bf r}) \} }{\delta {\bf X}({\bf r})} \right|_{{\bf X}({\bf r})= {\bf X}_{0}({\bf r} ; \theta(t))}, {\bf p}\{ {\bf X}_{0}({\bf r} ; \theta(t)), {\bf r}, t \} \right],
\label{seqX2}
\end{align}
which is correct up to $O(|| {\bf p} ||)$.  By denoting ${\bf p}(\theta, {\bf r}, t) = {\bf p}\{ {\bf X}_{0}({\bf r} ; \theta), {\bf r}, t \}$ and introducing a {\em phase sensitivity function}
\begin{align}
	{\bf Q}({\bf r} ; \theta)
	= \left. \frac{\delta \Theta\{ {\bf X}({\bf r}) \} }{\delta {\bf X}({\bf r})} \right|_{{\bf X}({\bf r}) = {\bf X}_{0}({\bf r} ; \theta)},
\label{seq19}
\end{align}
the reduced phase equation~(\ref{seqX2}) can be concisely written as
\begin{align}
\dot{\theta}(t) = \omega + \left[ {\bf Q}({\bf r} ; \theta), {\bf p}(\theta, {\bf r}, t)\right]
\label{seqP2}
\end{align}
at the lowest order in the perturbation.  The function ${\bf Q}({\bf r}; \theta)$ is the (functional) gradient of the isochron estimated at ${\bf X}({\bf r}) = {\bf X}_{0}({\bf r} ; \theta)$ on the limit-cycle solution $\chi$ and plays the key role in the present theory.

\subsection{Linear theory around the limit-cycle solution}

Though we have formally developed a geometric theory of phase reduction for the RD system, it is impossible to obtain $\Theta\{ {\bf X}({\bf r}) \}$ explicitly for general RD systems.
However, only the phase sensitivity function ${\bf Q}({\bf r} ; \theta)$ is actually necessary to write down the lowest-order phase equation~(\ref{seqP2}) for the weakly perturbed RD systems Eq.~(\ref{seqP1}).
We thus try to derive the equation for ${\bf Q}({\bf r} ; \theta)$ as in the ODE case, focusing only on the vicinity of the limit cycle $\chi$.

We first note that the phase function $\Theta({\bf X})$ for the ODE can be linearly approximated as
$ \Theta({\bf X}) \simeq \theta + {\bf Z}(\theta) \cdot ( {\bf X} - {\bf X}_{0}(\theta) )$
for the oscillator state ${\bf X}$ near ${\bf X}_{0}(\theta)$ on $\chi$ using the phase sensitivity function ${\bf Z}(\theta)$ defined in Eq.~(\ref{seq8})  (see Appendix A).
In a similar spirit, we make an ansatz that the phase $\Theta\{ {\bf X}({\bf r}) \}$ of a state ${\bf X}({\bf r})$ of the RD system near the state ${\bf X}_{0}({\bf r} ; \theta)$ on the limit cycle $\chi$ can be linearly approximated as
\begin{align}
	\Theta\{ {\bf X}({\bf r}) \} \simeq \theta + \left[ {\bf Q}({\bf r} ; \theta),\ {\bf X}({\bf r}) - {\bf X}_{0}({\bf r}; \theta) \right],
\label{seq20}
\end{align}
and examine whether this ansatz is reasonable.
When ${\bf X}({\bf r})$ is simply a state on the limit cycle $\chi$ with phase $\theta$, i.e., ${\bf X}({\bf r}) = {\bf X}_{0}({\bf r} ; \theta)$, Eq.~(\ref{seq20}) gives $\Theta\{ {\bf X}_{0}({\bf r} ; \theta) \} = \theta$.
If Eq.~(\ref{seq20}) is furthermore valid for arbitrary states $\{ {\bf X}({\bf r}) \}$ sufficiently close to the unperturbed state ${\bf X}_{0}({\bf r} ; \theta)$ on $\chi$, the function ${\bf Q}({\bf r} ; \theta)$ will play the role of the phase sensitivity function for the RD system.
This actually holds by choosing the function ${\bf Q}({\bf r} ; \theta)$ appropriately.
See Fig.~\ref{figS0RD}(b) for a schematic illustration.

We now derive the equation for ${\bf Q}({\bf r} ; \theta)$ by generalizing the argument in Refs.~\cite{ermentrout10,brown04} for the phase sensitivity function ${\bf Z}(\theta)$ of limit cycles described by ODEs (see also Appendix A).
At $t=0$, we prepare a initial state ${\bf X}_{0}({\bf r} ; \theta = 0) = {\bf X}_{0}({\bf r}, t=0)$ on the limit cycle $\chi$ with phase $\theta = 0$, and a slightly perturbed initial state
\begin{align}
{\bf X}({\bf r}, t=0) = {\bf X}_{0}({\bf r} ; \theta = 0) + {\bf y}({\bf r}, t=0)
\end{align}
near ${\bf X}_{0}({\bf r} ; \theta = 0)$, where ${\bf y}({\bf r}, t=0)$ is a small spatial perturbation given to ${\bf X}_{0}({\bf r} ; \theta = 0)$.
We evolve these two states without applying further perturbations.
Linearized equation for ${\bf y}({\bf r}, t) = {\bf X}({\bf r}, t) - {\bf X}_{0}({\bf r} ; \theta(t))$ can be obtained from Eq.~(\ref{seq14}) as
\begin{align}
\frac{\partial}{\partial t} {\bf y}({\bf r}, t) = {\rm J}( \theta(t) ) {\bf y}({\bf r}, t) + {\rm D} \nabla^{2} {\bf y}({\bf r}, t),
\label{seq23}
\end{align}
where ${\rm J}(\theta) = {\rm J}( {\bf X}_{0}({\bf r} ; \theta) )$ is a Jacobi matrix of ${\bf F}$ estimated at ${\bf X} = {\bf X}_{0}({\bf r} ; \theta)$ on $\chi$.
From Eq.~(\ref{seq20}), the phase of the unperturbed state ${\bf X}_{0}({\bf r} ; \theta(t))$ is $\theta(t) = \Theta\{ {\bf X}_{0}({\bf r} ; \theta(t) ) \} = \omega t$, and the phase of the perturbed state ${\bf X}({\bf r}, t )$ is given by
\begin{align}
\theta'(t) = \Theta\{ {\bf X}({\bf r}, t ) \} = \Theta\{ {\bf X}_{0}({\bf r} ; \theta(t)) + {\bf y}({\bf r}, t) \} \simeq \theta(t) + [ {\bf Q}({\bf r} ; \theta(t)),\ {\bf y}({\bf r}, t) ]
\end{align}
under the linear approximation.
This $\theta'(t)$ should also increase with a constant frequency $\omega$ by the definition of the isochron, i.e., $d\theta'(t)/dt = \omega$, because no perturbation is given after $t=0$.
Therefore, the following equation should hold:
\begin{align}
0
&= \frac{d}{d t} \big( \theta'(t) - \theta(t) \big)
=
\frac{\partial}{\partial t} \left[ {\bf Q}({\bf r} ; \theta(t)),\ {\bf y}({\bf r}, t) \right]
\cr
&=
\left[ \frac{\partial}{\partial t} {\bf Q}({\bf r} ; \theta(t)),\ {\bf y}({\bf r}, t) \right]
+
\left[ {\bf Q}({\bf r} ; \theta(t)),\ \frac{\partial}{\partial t} {\bf y}({\bf r}, t) \right].
\label{seq25}
\end{align}
Using Eq.~(\ref{seq23}), the last term can be transformed as
\begin{align}
\left[ {\bf Q}({\bf r} ; \theta(t)),\ \frac{\partial}{\partial t} {\bf y}({\bf r}, t) \right]
&=
\left[ {\bf Q}({\bf r} ; \theta(t)),\ {\rm J}( \theta(t) ) {\bf y}({\bf r}, t) + {\rm D} \nabla^{2} {\bf y}({\bf r}, t) \right] \cr
&=
\left[ {\rm J}( \theta(t) )^{\dagger} {\bf Q}({\bf r} ; \theta(t)) + {\rm D}^{\dagger} \nabla^{2} {\bf Q}({\bf r} ; \theta(t)),\ {\bf y}({\bf r}, t) \right],
\end{align}
where, as usual,  partial integration is performed assuming surface terms to vanish or cancel, and $\dagger$ denotes matrix transpose.
Equation~(\ref{seq25}) now yields
\begin{align}
\left[ 
\frac{\partial}{\partial t} {\bf Q}({\bf r} ; \theta(t)) + {\rm J}( \theta(t) )^{\dagger} {\bf Q}({\bf r} ; \theta(t)) + {\rm D}^{\dagger} \nabla^{2} {\bf Q}({\bf r} ; \theta(t)),\ {\bf y}({\bf r}, t)
\right] = 0,
\end{align}
which should hold for any ${\bf y}({\bf r}, t)$ evolving from arbitrary ${\bf y}({\bf r}, t=0)$.  Therefore, ${\bf Q}({\bf r} ; \theta(t))$ should satisfy
the following adjoint equation:
\begin{align}
\frac{\partial}{\partial t} {\bf Q}({\bf r} ; \theta(t)) = - {\rm J}( \theta(t) )^{\dagger} {\bf Q}({\bf r} ; \theta(t)) - {\rm D}^{\dagger} \nabla^{2} {\bf Q}({\bf r} ; \theta(t)),
\end{align}
or equivalently, by using the relation $\partial / \partial t = \omega \partial / \partial \theta$ ($\theta = \omega t$),
\begin{align}
\omega \frac{\partial}{\partial \theta} {\bf Q}({\bf r} ; \theta) = - {\rm J}( \theta )^{\dagger} {\bf Q}({\bf r} ; \theta) - {\rm D}^{\dagger} \nabla^{2} {\bf Q}({\bf r} ; \theta),
\label{seq29}
\end{align}
which we presented as Eq.~(3) in Section II.

Since this adjoint equation is linear, we also need to normalize ${\bf Q}({\bf r} ; \theta)$ appropriately.
As in the ODE case, the normalization condition for ${\bf Q}({\bf r} ; \theta)$ can be obtained by differentiating the identity $\theta(t) = \Theta\{ {\bf X}_{0}({\bf r} ; \theta(t)) \} = \omega t $ by $t$ as
\begin{align}
\omega
&= \frac{d}{dt} \theta(t) = \frac{d}{d t} \Theta\{ {\bf X}_{0}({\bf r} ; \theta(t)) \}
= \left[ \left. \frac{\delta \Theta\{ {\bf X}({\bf r}) \} }{\delta {\bf X}({\bf r})} \right|_{{\bf X}({\bf r}) = {\bf X}_{0}({\bf r} ; \theta(t))},\ \frac{\partial}{\partial t} {\bf X}_{0}({\bf r} ; \theta(t)) \right] \cr
&= \left[ {\bf Q}({\bf r} ; \theta(t)),\ {\bf F}({\bf X}_{0}({\bf r} ; \theta(t)), {\bf r}) + {\rm D} \nabla^{2} {\bf X}_{0}({\bf r} ; \theta(t)) \right].
\end{align}
Therefore, the following normalization condition should be satisfied:
\begin{align}
\omega = \left[ {\bf Q}({\bf r} ; \theta),\ {\bf F}({\bf X}_{0}({\bf r} ; \theta), {\bf r}) + {\rm D} \nabla^{2} {\bf X}_{0}({\bf r} ; \theta) \right].
\label{seq30}
\end{align}
Note that this condition can also be expressed, using again the relation $\partial / \partial t = \omega \partial / \partial \theta$, as
\begin{align}
\left[ {\bf Q}({\bf r} ; \theta),\  \frac{\partial}{\partial \theta} {\bf X}_{0}({\bf r} ; \theta) \right] = 1,
\label{seqconst}
\end{align}
which yields the normalization condition Eq.~(\ref{eq:normalization}) given in Section II.

Thus, if the function ${\bf Q}({\bf r} ; \theta)$ is the $2\pi$-periodic solution to Eq.~(\ref{seq29}) with the constraint (\ref{seqconst}), the approximate phase function Eq.~(\ref{seq20}) will satisfy the desired condition Eq.~(\ref{seq25}) in the vicinity of $\chi$, and ${\bf Q}({\bf r} ; \theta)$ will play the role of the phase sensitivity function of the limit-cycle solution $\chi$ of the RD system.
Note that Eqs.~(\ref{seq29}) and (\ref{seqconst}) are straightforward generalizations of the conventional adjoint method for the ODE, Eq.~(\ref{seq12}) and Eq.~(\ref{seq13}).
Generally, the function ${\bf Q}({\bf r} ; \theta)$ should be calculated numerically by solving the adjoint Eq.~(\ref{seq29}) with Eq.~(\ref{seqconst}).
In numerical calculations, it is useful to integrate the adjoint equation~(\ref{seq29}) backward in time to avoid numerical overflow with occasional normalization by Eq.~(\ref{seqconst}), as we explained in Section~III~A.

Once we obtain the frequency $\omega$ and the phase sensitivity function ${\bf Q}({\bf r} ; \theta)$, we can write down the approximate phase equation (\ref{seqP2}) for a slightly perturbed RD system described by Eq.~(\ref{seqP1}).
Note that the infinite-dimensional RD system subjected to weak perturbations is reduced to a single one-dimensional phase equation, which drastically simplifies the analysis of weakly perturbed rhythmic spatiotemporal patterns.
As a simple example of this phase reduction theory for RD systems, we analyzed synchronization dynamics of a pair of coupled RD systems exhibiting rhythmic patterns
in Section III, i.e., the circulating pulses, oscillating spots, target waves, and rotating spirals of the FitzHugh-Nagumo model.

\end{document}